\pdfoutput=1
\documentclass[a4paper,11pt]{article}

\usepackage{a4wide}

\usepackage{ifpdf}
\ifpdf
\fi

\ifpdf \def\drvr{pdftex} \else \def\drvr{dvips} \fi

\usepackage[utf8]{inputenc}
\usepackage[T1]{fontenc}
\usepackage[british]{babel}

\usepackage[table,\drvr]{xcolor}

\usepackage{lmodern}

\usepackage{xspace}

\usepackage{booktabs}

\usepackage{amsmath}
\usepackage{amssymb}
\usepackage{amsfonts}
\usepackage{bbm}
\usepackage{latexsym}

\usepackage{authblk}

\usepackage{listings}
\usepackage[\drvr]{hyperref}
\hypersetup{a4paper,setpagesize%
,colorlinks=false%
,unicode=true%
,breaklinks=true%
,pdfstartview={FitH}%
,pdfauthor={Moritz Karbach, Gerhard Raven, Manuel Schiller}%
,pdftitle={Complex error function and related integrals in neutral meson mixing}%
,pdfsubject={Complex error function}%
,pdfkeywords={mixing;convolution;integral}%
}

\usepackage[\drvr]{thumbpdf}

\ifpdf
\usepackage[protrusion=true,expansion=true,tracking]{microtype}
\fi

\usepackage[natbib=true,sorting=none]{biblatex}
\bibliography{main}

\title{Decay time integrals in neutral meson mixing and their efficient
evaluation}
\author[1]{Till Moritz Karbach}
\author[2]{Gerhard Raven}
\author[2]{Manuel Schiller}
\affil[1]{CERN, Switzerland}
\affil[2]{NIKHEF, The Netherlands}

\begin{document}
\maketitle

\begin{abstract}
  \noindent
  In neutral meson mixing, a certain class of convolution integrals is required
  whose solution involves the error function $\mathrm{erf}(z)$ of a complex
  argument $z$. We show the the general shape of the analytic solution of
  these integrals, and give expressions which allow the normalisation of these
  expressions for use in probability density functions. Furthermore, we derive
  expressions which allow a (decay time) acceptance to be included in these
  integrals, or allow the calculation of moments.

  \noindent
  We also describe the implementation of numerical routines which allow the
  numerical evaluation of $w(z)=e^{-z^2}(1-\mathrm{erf}(-iz))$, sometimes also
  called Faddeeva function, in {\tt C++}. These new routines improve over the old
  CERNLIB routine(s) {\tt WWERF}/{\tt CWERF} in terms of both speed and
  accuracy. These new routines are part of the {\sc RooFit} package, and have
  been distributed with it since {\sc ROOT} version 5.34/08.
\end{abstract}
\clearpage

\renewcommand{\d}[1]{\ensuremath{{\rm d}#1}\xspace}

\renewcommand{\thefootnote}{\arabic{footnote}}
\setcounter{footnote}{0}

\tableofcontents
\clearpage

\pagestyle{plain} % restore page numbers for the main text
\setcounter{page}{1}
\pagenumbering{arabic}

% $Id: introduction.tex 39473 2013-07-19 03:25:20Z mschille $

%The reason for this interest is
%due to the work on physics analyses the authors are performing which need to
%describe $B$ and/or $D$ mixing phenomena.

\section{Introduction}
\label{sec:Introduction}

When dealing with a time-dependent analysis of neutral mesons, one encounters
the effect of meson mixing, leading to decay rate equations of the form
\begin{equation}
	\frac{\d{\Gamma_{\rm theo}(t)}}{\d{t}} \sim e^{-\Gamma t}\left(
	A\cosh(\Delta\Gamma t/2) + B\sinh(\Delta\Gamma t/2) +
	C\cos(\Delta mt) + D\sin(\Delta mt) \right)\,,
\end{equation}
for $t > 0$ with real coefficients $A$, $B$, $C$, and $D$, where $\Gamma$ is the
average width of the two meson mass eigenstates and $\Delta\Gamma$ and $\Delta
m$ are the width and mass difference between the mass eigenstates,
respectively.
Usually, the decay time resolution of the detector is finite, so this has to be
convoluted with a resolution model, e.g. a Gaussian, to give the experimentally
observable decay rate
\begin{equation}
	\frac{\d{\Gamma}_{\rm exp}(t)}{\d{t}} = \int_{-\infty}^{+\infty} \d{t'}\,\theta(t')
	\frac{\d{\Gamma}_{\rm theo}(t')}{\d{t'}}\,G(t-t',\mu,\sigma)\,,
	\label{eq:convolution}
\end{equation}
where
\begin{equation}
	G(t-t',\mu,\sigma)=\frac{1}{\sqrt{2\pi\sigma^2}}e^{-\frac{(t-t'-\mu)^2}{2\sigma^2}}\,,
\end{equation}
and $\theta(t')$ is the Heavyside (step) function.
The parameter $\mu$ represents a potential bias in the reconstructed decay time, and $\sigma$
is the decay time resolution. Due to the linearity of Eq.~\ref{eq:convolution}, it is
trivial to extend to a more realistic, multi-Gaussian resolution function.
In addition to a finite time resolution, most detectors show detection, reconstruction and/or trigger
efficiency variations as a function of decay time, which need to be modeled
by an \emph{acceptance function} $a(t)$. The final acceptance-corrected decay
rate equation becomes:
\begin{equation}
	\frac{\d{\Gamma}^{\rm acc}_{\rm exp}(t)}{\d{t}} = \frac{\d{\Gamma}_{\rm exp}(t)}{\d{t}}\,a(t)\,.
	\label{eq:accconvolution}
\end{equation}
To use either Eq.~\ref{eq:convolution} or Eq.~\ref{eq:accconvolution} as building blocks
for a probability density function, the equations need to be normalised by
dividing by their integral over
the observable (the decay time $t$ in this case).

The purpose of this note is to demonstrate how to solve the relevant
integrals analytically, and to collect the resulting expressions for
future reference. In addition, a numerical implementation of the the 
\emph{Faddeeva} function is provided, which is, as will be shown, an essential 
part of the expressions.

This document is organised as follows.
Section~\ref{sec:definitions} defines the error function and some related
functions and reviews their properties. The convolution integral in 
Eq.~\ref{eq:convolution} and its normalisation integral are solved analytically in
Sect.~\ref{sec:convolution}. Section~\ref{sec:accconvolution} deals with the
normalisation of Eq.~\ref{eq:accconvolution} and the calculation of moments of
Eq.~\ref{eq:convolution}. Section~\ref{sec:implementation} discusses a computer
program to compute the \emph{Faddeeva} function numerically.

\section{Definitions\label{sec:definitions}}
\subsection{Error Function}
The \emph{error function} is defined for real argument $x$ as
\begin{equation}
{\rm erf}(x)=\frac{2}{\sqrt{\pi}}\int_0^x \d{t}\, e^{-t^2}.
\end{equation}
It is an odd function, i.e.
\begin{equation}
{\rm erf}(-x) = -{\rm erf}(x).
\end{equation}
This definition can be continued into the complex plane with a complex argument
$z$ taking the place of $x$. The resulting function is analytic over the
entire complex plane, and in general takes complex values. There is an
additional symmetry in the complex plane:
\begin{equation}
{\rm erf}(\overline{z})=\overline{{\rm erf}(z)}\,.
\end{equation}
The integral and derivative of the error function are given by
\begin{equation}
\int \d{z}\,{\rm erf}(z)=z\,{\rm erf}(z)+\frac{e^{-z^2}}{\sqrt{\pi}}\,.
\end{equation}

\subsection{Complementary Error Function}
The \emph{complementary error function} is defined as
\begin{equation}
{\rm erfc}(x) = 1-{\rm erf}(x)\,.
\end{equation}
Its symmetry property is
\begin{equation}
{\rm erfc}(-x)=2-{\rm erfc}(x)\,.
\end{equation}
The continuation into the complex plane yields an analytic
function with the same symmetry with respect to complex conjugation as the
error function itself:
\begin{equation}
{\rm erfc}(\overline{z})=\overline{{\rm erfc}(z)}\,.
\end{equation}
The integral of the complementary error function is given by
\begin{equation}
\int \d{z}\,{\rm erfc}(z)=z\,{\rm erfc}(z)-\frac{e^{-z^2}}{\sqrt{\pi}}\,.
\label{eq:interfc}
\end{equation}

\subsection{Faddeeva Function}
The Faddeeva function $w(z)$ is closely related to the error function, it is defined as
\begin{equation}
w(z)=e^{-z^2}{\rm erfc}(-iz)\,.
\end{equation}
This function has the symmetries
\begin{equation}
w(-x+iy)=\overline{w(x+iy)}\,,
\qquad
w(x-iy)=2e^{-z^2}-w(x+iy)\,.
\label{eq:wsymmetry}
\end{equation}
Its derivative is given by
\begin{equation}
\frac{\d{}}{\d{z}}w(z)=\frac{2i}{\sqrt{\pi}} -2z\,w(z)\,.
\end{equation}

\section{Neutral Meson Mixing in the Presence of Decay Time Resolution
\label{sec:convolution}}
In Eq.~\ref{eq:convolution}, there are three intrinsic time
scales: the decay time $1/\Gamma > 0$, the oscillation period $1/\Delta m > 0$
and the time resolution $\sigma > 0$\footnote{One may argue that there is a
fourth scale, $\Delta\Gamma$, entering the problem. However, this can trivially
be reduced to two different lifetimes $\Gamma_1$ and $\Gamma_2$ for the two
mass eigenstates of the problem. Each mass eigenstate is then treated
separately, leaving only three time scales.}. The numerical stability of the
expressions we are about to derive depends on the relative orders of magnitude
of $1/\Gamma$, $1/\Delta m$ and $\sigma$. We consider the solution of the
convolution integral in Eq.~\ref{eq:convolution} in three cases:
\begin{enumerate}
\item the general case,
\item $\min(1/\Gamma,1/\Delta m) \gg \sigma$ (i.e.~the detector resolution is much better than either lifetime or oscillation frequency demand), and
\item $\min(\sigma,1/\Delta m) \gg 1/\Gamma$ (i.e.~the decay is so fast that all decaying particles can be said to decay at the same time).
\end{enumerate}
The general case is discussed within the main text, the two special cases 2 and
3 have been moved to Appendix \ref{sec:specconvolution}, as the matter is dry
enough as is.

\subsection{General Case}
Since $\sin(\Delta m t) = \Im\left( e^{i\Delta m
t}\right)$ and $\cos(\Delta m t ) = \Re\left( e^{ i \Delta m t } \right)$, and
both the $\cosh$ and $\sinh$ terms in Eq.~\ref{eq:convolution} can be
written as the sum and difference of exponentials, it is sufficient to consider the following
convolution:
\begin{align}
f(t;\Gamma,\Delta m,\sigma,\mu)
&= \frac{1}{\sqrt{2\pi\sigma^2}}
	\int_0^{+\infty}\d{t'}\,e^{-(\Gamma -i\Delta m)t'}
	e^{-\frac{(t-t'-\mu)^2}{2\sigma^2}}
	\label{eq:mastereq}\\
&= \frac{1}{\sqrt{\pi}}\int_0^{+\infty}\d{y}\,e^{-(x-y)^2-2zy}\,,
\end{align}
where we have substituted $z = (\Gamma - i \Delta m ) \sigma/\sqrt{2}$, $x =
\frac{t-\mu}{\sqrt{2}\sigma}$ and $y=\frac{t'}{\sqrt{2}\sigma}$.
Completing the square in the exponent and
absorbing the shift in the boundaries of the integral, we find:
\begin{align}
f(x;z)
&= e^{-x^2+(z-x)^2} \frac{1}{\sqrt{\pi}} \int_0^{+\infty}\,\d{y}\,e^{-\left(y+(z-x) \right)^2} \\
&= e^{-x^2+(z-x)^2} \frac{1}{\sqrt{\pi}} \int_{z-x}^{+\infty}\,\d{y}\,e^{-y^2 } \\
&= e^{-x^2+(z-x)^2} \; \frac{1}{2}\mathrm{erfc}\left(z-x \right)\,.
\end{align}
Using the Faddeeva function $w(z)$, this can be written as
\begin{equation}
f(x;z) = \frac{1}{2} e^{-x^2} w\left( i \left(z-x\right)\right)\,.
\end{equation}
The corresponding normalisation integral is given by:
\begin{equation}
    I_0(t_1,t_2;\Gamma,\Delta m,\sigma,\mu) = \int_{t_1}^{t_2}\d{t}\,f(t;\Gamma,\Delta m,
\sigma,\mu) = \sigma\sqrt{2} \int_{x_1}^{x_2}\d{x}\,f(x;z) = \frac{\sigma}{\sqrt{2}} \hat{I}_0(x_1,x_2;z)\,,
\end{equation}
where the latter $\hat{I}_0(x_1,x_2;z)$ is defined as
\begin{align}
   \hat{I}_0(x_1,x_2;z) &\equiv
    \int_{x_1}^{x_2} \d{x}\,e^{-x^2} w\left( i(z-x) \right)\\
   &=\frac{1}{2z}\left[ \mathrm{erf}\left(x\right) -e^{-x^2}
w\left( i\left(z-x\right) \right) \right]_{x_1}^{x_2}\,,
\end{align}
where we have used Eq.~\ref{eq:interfc}.

\section{Calculating Moments and Including the Effect of an Acceptance Function\label{sec:accconvolution}}
To describe a non-trivial decay time acceptance, one generally approximates
$a(t)$ in some way, e.g.~by piecewise constant or linear functions, or by
piecewise polynomials such as splines. In these cases, it is sufficient to restrict the 
problem to functions $a(t)$ which are of the form  $a(t)=\sum_k a_kt^k$.
To normalise $\sum_k a_kt^k\cdot f(t;\Gamma,\Delta m,\sigma,\mu)$, one needs to
compute the integrals
\begin{equation}
I_k(t_1,t_2;\Gamma,\Delta m,\sigma,\mu)=\int_{t_1}^{t_2} t^k f(t;\Gamma,\Delta
m,\sigma,\mu)\,\d{t}\,.
\label{eq:t-integral}
\end{equation}
These integrals also define the moments $m_k$,
\begin{equation}
m_k=\frac{\int \d{t}\,t^k\, f(t)}{\int \d{t}\,f(t)} = \frac{I_k}{I_0}\,.
\end{equation}
When computing these integrals we again consider the three cases from the last
section, where the two special cases 2 and 3 from the last section can be found
in Appendix \ref{sec:specaccconvolution}.

\subsection{General Case}
In the general case, it is again useful to go to the reduced coordinates $x$
and $z$ defined in the last section. The integral in Eq.~\ref{eq:t-integral}
then becomes:
\begin{align}
I_k(x_1,x_2;z)
    &=\int_{x_1}^{x_2}\left(\sqrt{2}\sigma x+\mu\right)^k\,f(x;z)\,\sqrt{2}\sigma\,\d{x} \\
    &=\sqrt{2}\sigma \sum_{n=0}^{k}{k \choose n}\left(\sqrt{2}\sigma \right)^n \mu^{k-n} \int_{x_1}^{x_2}\d{x}\, x^{n}\,f(x;z) \\
    &\equiv\frac{\sigma}{\sqrt{2}}\sum_{n=0}^{k}{k \choose n}\left(\sqrt{2}\sigma \right)^n\,\mu^{k-n}\,\hat{I}_n(x_1,x_2;z)\,.
\end{align}
The required integrals are thus
\begin{equation*}
    \hat{I}_n(x_1,x_2;z) = \int_{x_1}^{x_2} \d{x}\, x^n\, e^{-x^2}\, w\left( i (z-x)\right)\,.
\end{equation*}
They can be computed using the following method:
\begin{align*}
    \hat{I}_n(x_1,x_2;z)&\equiv
    	\int_{x_1}^{x_2}\,\d{x}\,x^n e^{-x^2} w\left( i (z-x)\right)\\ 
    &=\frac{1}{2^n}\left.\frac{\d{}^n}{\d{\lambda}^n}\right|_{\lambda=0 }
    	\int_{x_1}^{x_2}\d{x}\,e^{2\lambda x} e^{-x^2} w\left( i(z-x)\right)\\
	&\equiv \frac{1}{2^n} \left.\frac{\d{}^n}{\d{\lambda}^n}\right|_{\lambda=0 } \hat{I}(x_1,x_2;z,\lambda)\,.
\end{align*}
Thus, we rewrite the term $x^n$ as the slightly more complicated expression
\begin{equation*}
x^n \equiv \left.\frac{1}{2^n}\frac{\d{}^n}{\d{\lambda^n}}\right|_{\lambda=0}e^{2\lambda x}\,,
\end{equation*}
to obtain an expression where the integration and the derivative with respect
to $\lambda$ commute. This facilitates the treatment of the integral
enormously. Once again we complete the square, and shift the integrand to
obtain:
\begin{align*}
    \hat{I}(x_1,x_2;z,\lambda) &=
		\int_{x_1}^{x_2}\d{x}\,e^{2\lambda x} e^{-x^2} w\left( i(z-x)\right)\\
    &= e^{\lambda^2}\int_{x_1-\lambda}^{x_2-\lambda}\d{x}\,e^{-x^2} w\left(i\left(z-\lambda-x\right) \right)\\
    &= \frac{e^{\lambda^2}}{2(z-\lambda)}
    	\left[{\rm erf}\left( x \right) - e^{-x^2}w\left(i\left(z-\lambda-x \right) \right) \right]_{x_1-\lambda}^{x_2-\lambda}\\
    &\equiv K(z,\lambda) \left[ J(x_2;\lambda,z) - J(x_1;\lambda,z)  \right]\,,
\end{align*}
where
\begin{equation*}
K(\lambda,z)=\frac{e^{\lambda^2}}{z-\lambda}\,,\qquad 
J(x;\lambda,z)=\mathrm{erf}\left(x-\lambda\right)-e^{-\left(x-\lambda\right)^2}w\left(i\left(z-x\right) \right)\,. 
\end{equation*}
In order to simplify the  computation of the $\hat{I}_n$, we compute the $n^\mathrm{th}$ 
order derivatives at $\lambda=0$, $K_n(z)$ and $M_n(x;z)$, as follows:
\begin{center}\begin{tabular}{c c c}
	\toprule
    $n$
    & $K_n(z)$
	& $M_n(x;z)$ \\
    \midrule
   $0$   & $\frac{1}{2z}$                              & $\mathrm{erf}\left(x\right)-e^{-x^2}w\left( i \left(z-x\right)\right)$
\\ $1$   & $\frac{1}{2z^2}$                            & $2 e^{-x^2} \left[ - \sqrt{\frac{1}{\pi}}   -       x w\left(i(z-x)  \right) \right]$
\\ $2$   & $\frac{1}{z}\left(1 + \frac{1}{z^2}\right)$ & $2 e^{-x^2} \left[ - 2x  \sqrt{\frac{1}{\pi}}  - (2x^2-1) w\left(i(z-x)\right) \right]$
\\ $3$   & $\frac{3}{z^2}\left(1+\frac{1}{z^2}\right)$ & $4 e^{-x^2} \left[ - (2x^2-1)\sqrt{\frac{1}{\pi}} -x(2x^2-3)w\left(i(z-x)\right)   \right]$
\\ \bottomrule
\end{tabular}
\end{center}
The normalisation integrals in terms of $K_n(z)$ and $M_n(x;z)$
are thus
\begin{align*}
    \hat{I}_n(x_{1},x_{2}, z) &=
    	\left[\frac{1}{2^n}\left.\frac{\d{}^n}{\d{\lambda}^n}\right|_{\lambda=0} K(\lambda ,z) J(x;\lambda ,z)
		\right]^{x_{2}}_{x_{1}} \\
	&= \left[\frac{1}{2^n}\sum_{k=0}^n {n \choose k} K_k(z) M_{n-k}(x,z)\right]^{x_{2}}_{x_{1}}  \\
	&\equiv \frac{1}{2^n}\sum_{k=0}^n {n \choose k} K_k(z) M_{n-k}(x_1,x_2,z) \,,
\end{align*}
where we have defined the abbreviation $M_n(x_1,x_2;z)\equiv M_n(x_2;z)-M_n(x_1;z)$.
Given that the typical use requires the computation of the sum over several $\hat{I}_n$, e.g.
\begin{equation}
  N(x_1,x_2,z) = \sum_{k=0}^{n} a_k \hat{I}_k(x_1,x_2,z)\,,
  \label{eq:sumofintegrals}
\end{equation}
it is advantageous to reorder the implied double sum:
\begin{equation}
  N(x_1,x_2,z) = \sum_{i=0}^{n} \sum_{j=0}^i A_{ij} M_i(x_1,x_2;z) K_j(z)\,,
\end{equation}
where the matrix $A$ is defined by
\begin{equation}
A_{ij} \equiv  \left\{\begin{array}{ll}
    \frac{a_{i+j}}{2^{i+j}}\left( \begin{array}{c} i+j \\ j \end{array} \right)& \textrm{for }i+j\le n\\
	0 & \textrm{otherwise} \\
    \end{array}\right. \, .
\end{equation}
Now the dependence on the coefficients $a_k$  of Eq. \ref{eq:sumofintegrals} can be fully absorbed 
in the definition of the matrix $A$. 
For example, in case of $n = 3$, when written in a vector notation, this results in:
\begin{equation}
  N(x_1,x_2,z) = 
       \left(\begin{array}{c} M_0(x_1,x_2;z) \\ M_1(x_1,x_2;z) \\ M_2(x_1,x_2;z) \\ M_3(x_1,x_2;z) \end{array}\right)
\left(\begin{array}{cccc}
                   a_0  & \frac{a_1}{2} & \frac{a_2}{4}  &  \frac{a_3}{8}
               \\  \frac{a_1}{2}  & \frac{a_2}{2} & \frac{3a_3}{8}  &  0
               \\  \frac{a_2}{4}  & \frac{3a_3}{8} & 0  &  0
               \\  \frac{a_3}{8}  & 0 & 0  &  0
               \end{array}\right) 
       \left(\begin{array}{c} K_0(z) \\ K_1(z) \\ K_2(z) \\ K_3(z) \end{array} \right)\,.
\end{equation}

\section{Evaluation of the Faddeeva Function in software\label{sec:implementation}}
\subsection{Implementation}
To implement the Faddeeva function, we largely follow the ideas in
\cite{Abrarov20111894}, which we will sketch briefly below. Our code is also
included in Appendix \ref{sec:code}.

The aim is to implement a full precision version which yields results that
are accurate to within a few times the machine precision of a {\tt C++} {\tt
double} (64 bits, about $2\cdot10^{-16})$, and a faster version which is
accurate to a few times the machine precision of a {\tt C++} {\tt float} (32
bits, about $1\cdot10^{-7}$).  We start from an alternative formulation of
the Faddeeva function by representing it with a Fourier-style integral:
\begin{equation}
w(z)=\frac{1}{\sqrt{\pi}}\int_0^\infty\d{\tau}\,e^{-\frac{\tau^2}{4}}e^{i\tau z}\,.
\label{eq:wzfourier}
\end{equation}
The idea is now to approximate the term $e^{-\tau^2/4}$ as a Fourier
series 
\begin{equation}
    e^{-\frac{\tau^2}{4}}\approx\sum_{n=0}^{N} \frac{a_n}{2}\left(
	e^{\frac{in\pi}{\tau_m}\tau}+
    e^{-\frac{in\pi}{\tau_m}\tau}\right)-\frac{a_0}{2}\,,
    \qquad a_n\approx\frac{2\sqrt{\pi}}{\tau_m}e^{-\frac{n^2\pi^2}{\tau_m^2}}\,,
    \label{eq:expfourier}
\end{equation}
in the interval $-\tau_m\le\tau\le\tau_m$, where the $a_n$
are the with Fourier coefficients.
The resulting equation is
\begin{align}
w(z)&\approx\frac{1}{\sqrt{\pi}}\int_0^\infty\d{\tau}\,
    \left(\sum_{n=0}^{N} \frac{a_n}{2}\left(
	e^{\frac{in\pi}{\tau_m}\tau}+
	e^{-\frac{in\pi}{\tau_m}\tau}\right)-\frac{a_0}{2}\right)
	e^{i\tau z} \\
&=\frac{i}{2\sqrt{\pi}} \left(
    \sum_{n=0}^{N} a_n\tau_m\left(
	\frac{1-e^{i(n\pi+\tau_mz)}}{n\pi+\tau_mz}-
    \frac{1-e^{i(-n\pi+\tau_mz)}}{n\pi-\tau_mz}\right)-a_0\frac{1-e^{i\tau_mz}}{z}\right)\,.
    \label{eq:implmaster}
\end{align}
In the following we discuss how
\begin{enumerate}
\item to choose the integration cutoff $\tau_m$,
\item to choose $N$, and
\item the singularities at $z_n=\pm\frac{n\pi}{\tau_m}$ in Eq.~\ref{eq:implmaster} can be treated.
\end{enumerate}
The choice of $\tau_m$ is easiest: Since $\tau_m$ cuts off the
integral in Eq.~\ref{eq:wzfourier}, one needs to ensure that the portion of
the integral that is neglected is sufficiently small. To obtain {\tt double}
({\tt float}) precision, $e^{-\tau_m^2/4}$ should be on the order of
the machine precision of these data types, i.e. around $2\cdot10^{-16}$
($1\cdot10^{-7}$). This leads to the choices of $\tau_m=12$ for a full
precision version of the routine, and $\tau_m=8$ for a faster version with
reduced precision.

Next, we chose $N$. It has to be large enough that the Fourier series
in Eq.~\ref{eq:expfourier} is a good approximation. This is the case when
the highest Fourier coefficient is smaller than the machine precision of the
data type in question. For the full precision version, this means $N=23$, 
for the fast version with reduced precision it means $N=10$.

Finally, the singularities in Eq.~\ref{eq:implmaster} at $z_n=\pm\frac{n\pi}{\tau_m}$
are handled by using Taylor expansions of $w(z)$ in a tiny disc
$|z-z_n|<3\cdot10^{-3}$ around the singularities. To achieve the required
precision, one has to take into account terms up to the fifth (second) order
in $(z-z_n)$ for the slow (fast) version of the routine.
Outside the discs around the singularities, the code thus uses 
Eq.~\ref{eq:implmaster} for $\Re(z),\Im(z) \ge 0$, i.e. in the first quadrant of
the complex plane. For arguments $z$ outside the first quadrant of the
complex plane, the symmetries of the Faddeeva function 
(Eq.~\ref{eq:wsymmetry}) can be used. Thus, only $N+1$ Taylor expansions
of $w(z)$ need to be saved (and not $2N+1$), and the numerical instability
of $w(z)$ for $\Im(z)\ll0$ due to its divergent nature in this regime can
largely be avoided.

The code execution can also be optimised:
\begin{itemize}
\item The term $e^{in\pi}$ in Eq.~\ref{eq:implmaster} is a constant, $\pm1$,
    so there is no need to compute it.
\item The term $e^{i\tau_mz}$ in Eq.~\ref{eq:implmaster} depends only on $z$,
    so it can be precomputed at the beginning of the routine, avoiding a
    computationally expensive complex exponential inside the loop
    implementing the sum.
\item The subexpressions $n\pi$ and coefficients $a_n$ in 
	Eq.~\ref{eq:implmaster} can be precomputed before the code is compiled, and 
    provided by small lookup tables.
\item On the {\tt x86\_64} architecture, the GNU {\tt C++} compiler produces 
	suboptimal code for the complex
    exponentiation: Exponential and sine and cosine of a real argument are
    implemented in hardware and executed on the x87 unit of the CPU. Normal
    floating point operations like multiplication typically happen in
    another functional unit of the CPU, however. Both units have their
    separate floating point register sets, and moving values between the two
    involves a store to, and subsequent load from, the main memory (RAM). There
    are thus five of these load-store instruction pairs (real and imaginary
    part of the input argument to the complex error function, exponential of
    the real part, and sine and cosine of the imaginary part) which copy
    around input/output values.
    For this reason, the code includes a hand-coded inline assembly version
    of the complex exponential function, which saves at least one
    store-load instruction pair by computing the result entirely in the x87
    unit of the CPU. It also does away with the subroutine calls into
    the math library of the system. On all other systems, the code
    automatically uses the less optimal version in the math
    library.
\item The code contains two versions of the loop to compute the sum in 
	Eq.~\ref{eq:implmaster}: A na\"ive implementation, and one that is at least
    partially vectorisable with modern compilers. The latter will use SIMD
    instructions when available. The code chooses the version to use based
    on architecture-specific macros being defined during the compilation
    phase.
\item Due to the divergent nature of $w(z)$ for $\Im(z)\ll0$, the fast
    version of the routine also needs to use {\tt double} calculations
    internally to avoid loss of precision beyond the level we aim for based
    on our choices of $\tau_m$ and $N$.
\end{itemize}
The {\tt C++} code of our implementation is included in the Appendix. It has been
part of the \textsc{RooFit} package \cite{Verkerke:2003ir} since ROOT~\cite{ROOT} version 5.34/08.

\subsection{Performance}
In this subsection, we compare several packages to compute the Faddeeva, erf
and erfc functions for complex arguments. {\sc RooFit} is a fitting package 
in the ROOT framework which makes heavy use
of the Faddeeva function, so it makes sense to check the accuracy and speed
of various implementations. Specifically, we investigate:
\begin{itemize}
\item the original {\sc CERNLIB} {\tt WWERF} implementation
    \cite{CERNLIB} written in 1970 in FORTRAN77 (using an older algorithm),
\item the old code in {\sc RooFit} (before ROOT version 5.34/08); there is a
    slow version of the routine based on the CERNLIB implementation, ported
    to {\tt C++}, and a fast version, which is based on a 12.5 Megabyte  lookup
    table and interpolation in the rectangle defined by $|\Re(z)|<4$ and
    $-4\le\Im(z)\le6$ (which falls back on the slow version outside that area),
\item our code (in {\sc RooFit} since ROOT version 5.34/08), as
    described in the last subsection,
\item code based on the {\tt libcerf} library~\cite{libcerf} written in {\tt C}.
\end{itemize}
The {\tt libcerf} library provides special implementations for the erf and erfc
functions, the other packages use the following relations to define
these functions in terms of $w(z)$:
\[ \mathrm{erf}(z)=\left\{\begin{array}{l l}
	    1-e^{-z^2}w(iz) & \textrm{for }\Re(z)\ge0 \\
	    e^{-z^2}w(-iz) - 1& \textrm{otherwise}
    \end{array}\right.\,, \]
\[ \mathrm{erfc}(z)=\left\{\begin{array}{l l}
	    e^{-z^2}w(iz) & \textrm{for }\Re(z)\ge0 \\
	    2-e^{-z^2}w(-iz) & \textrm{otherwise}
    \end{array}\right.\,. \]

\subsubsection{Performance evaluation method}
To judge the numerical accuracy of these routines, we compare the results
of the implementations to
those obtained with the computer algebra system Maxima~\cite{maxima}.
In Maxima, one can calculate these functions using a special
``bigfloat'' floating point data type for which one can chose the length of
the mantissa at runtime. With 48 significant
decimal digits in the mantissa, the results of Maxima can be trusted to full
{\tt double} precision. More specifically, we calculate the absolute value
of the relative difference between the implementations under study and the
result obtained with Maxima, $\epsilon=|1-w(z)/w_{\mathrm{Maxima}}(z)|$.

To have an indication about the relative speed of the implementations
under study, we measure the number of CPU cycles needed for the execution of
the different routines using a hardware register incremented with each CPU
clock (on {\tt x86/x86\_64}, the {\tt TSC} register).

We consider two areas from which to choose $z$:
\begin{itemize}
    \item the ``big square'' $-8\le\Re(z),\Im(z)\le8$, in which we test
	$2^{16}$ random points distributed uniformly over that square, and
    \item the ``singularity'' areas where our algorithm has to switch to the
	Taylor expansions around $z_n=\frac{n\pi}{\tau_m}$; specifically,
	the area considered is
	$\mathrm{max}(\Re(|z-z_n|),\Im(|z-z_n|))<4\cdot10^{-3}$. For
	each of these $N$ squares, we test 1024 points distributed
	uniformly in that area.
\end{itemize}

\subsubsection{Results}
Tables~\ref{tab:perfsquare} and~\ref{tab:perfsing} show the performance
figures obtained on a typical laptop running a Linux system
with an Intel Core i7-2620M CPU running at 2.7 GHz. The compiler suite used
was the GNU compiler collection version 4.7.2 with optimisation options
``{\tt -O3 -ffast-math -fno-math-errno -mtune=native -mmmx -msse -msse2
-mssse3 -msse4.1 -msse4.2\\
-mavx}''.
We have also run the benchmarks on a different platform (Linux PowerPC G3,
a 32 bit machine with big endian byte order) to make sure that there are no
hidden portability pitfalls in our code. The accuracy is practically
unchanged for all implementations but {\tt libcerf}, which seems to produce
slightly different results for infinite arguments, arguments with very large
$|z|$, or arguments containing NaNs on the PowerPC machine.
Timings appear to be slightly different, but the general
trends are similar to those shown in Tables~\ref{tab:perfsquare}
and~\ref{tab:perfsing}.  Our code has been compiled and tested as
part of the ROOT releases on many different platforms giving confidence
that the code is quite portable, and delivers the same accuracy
independent of the particular IEEE754 floating point implementation used.

\begin{table}
\begin{center} \begin{tabular}{l c c c c c c}\toprule
    & {\sc CERNLIB} & {\tt libcerf} & \multicolumn{2}{c}{{\sc RooFit}
old} & \multicolumn{2}{c}{our code} \\
    $w(z)$     &&&(precise)&(fast)&(precise)&(fast) \\ \midrule
time $[$cycles$]$	& $2.4\cdot10^3$ & $9.8\cdot10^2$ & $2.8\cdot10^3$ & $2.1\cdot10^3$ & $6.8\cdot10^2$ & $5.3\cdot10^2$ \\
$\epsilon$		& $6.7\cdot10^{-15}$ & $1.6\cdot10^{-15}$ & $6.7\cdot10^{-15}$ & $6.2\cdot10^{- 8}$ & $6.1\cdot10^{-16}$ & $4.1\cdot10^{- 9}$ \\ 
$\epsilon_\mathrm{max}$	& $2.5\cdot10^{-12}$ & $8.4\cdot10^{-14}$ & $2.5\cdot10^{-12}$ & $5.1\cdot10^{- 5}$ & $8.4\cdot10^{-14}$ & $1.8\cdot10^{- 7}$ \\
\toprule $\mathrm{erf}(z)$ &&&&&& \\ \midrule
time $[$cycles$]$	& $2.5\cdot10^3$ & $1.2\cdot10^3$ & $2.9\cdot10^3$ & $2.0\cdot10^3$ & $7.9\cdot10^2$ & $6.4\cdot10^2$ \\
$\epsilon$		& $1.3\cdot10^{-14}$ & $1.4\cdot10^{-15}$ & $1.3\cdot10^{-14}$ & $6.4\cdot10^{- 9}$ & $1.1\cdot10^{-15}$ & $3.5\cdot10^{- 9}$ \\ 
$\epsilon_\mathrm{max}$	& $5.6\cdot10^{-11}$ & $8.4\cdot10^{-14}$ & $5.6\cdot10^{-11}$ & $6.0\cdot10^{- 6}$ & $8.4\cdot10^{-14}$ & $1.9\cdot10^{- 7}$ \\
\toprule $\mathrm{erfc}(z)$ &&&&&& \\ \midrule
time $[$cycles$]$	& $2.5\cdot10^3$ & $1.1\cdot10^3$ & $2.8\cdot10^3$ & $2.0\cdot10^3$ & $7.7\cdot10^2$ & $6.2\cdot10^2$ \\
$\epsilon$		& $7.1\cdot10^{-15}$ & $2.0\cdot10^{-15}$ & $7.1\cdot10^{-15}$ & $5.7\cdot10^{- 9}$ & $1.7\cdot10^{-15}$ & $4.0\cdot10^{- 9}$ \\ 
$\epsilon_\mathrm{max}$	& $2.6\cdot10^{-12}$ & $9.6\cdot10^{-14}$ & $2.6\cdot10^{-12}$ & $3.7\cdot10^{- 7}$ & $7.0\cdot10^{-14}$ & $1.9\cdot10^{- 7}$ \\
\bottomrule \end{tabular} \end{center}
\caption{Performance of the various implementations of $w(z)$,
    $\mathrm{erf}(z)$ and $\mathrm{erfc}(z)$ for $2^{16}$ values of $z$ from
    the ``big square'' region (see text). Time is measured in CPU cycles per
    evaluation (with variations of 5-10\% between different runs of the
    program on the same machine). $\epsilon$ is the average over the
    relative errors of all points tested, $\epsilon_{\mathrm{max}}$ is the
    maximum relative error seen.
}
\label{tab:perfsquare}
\end{table}

\begin{table}
\begin{center} \begin{tabular}{l c c c c c c}\toprule
    & {\sc CERNLIB} & {\tt libcerf} & \multicolumn{2}{c}{{\sc RooFit}
old} & \multicolumn{2}{c}{our code} \\
    $w(z)$     &&&(precise)&(fast)&(precise)&(fast) \\ \midrule
time $[$cycles$]$	& $3.0\cdot10^3$ & $1.1\cdot10^3$ & $3.4\cdot10^3$ & $1.5\cdot10^3$ & $5.9\cdot10^2$ & $5.0\cdot10^2$ \\
$\epsilon$		& $2.9\cdot10^{-12}$ & $3.4\cdot10^{-16}$ & $2.9\cdot10^{-12}$ & $1.2\cdot10^{- 7}$ & $4.1\cdot10^{-16}$ & $3.7\cdot10^{- 9}$ \\ 
$\epsilon_\mathrm{max}$	& $3.0\cdot10^{-12}$ & $1.6\cdot10^{-15}$ & $3.0\cdot10^{-12}$ & $3.8\cdot10^{- 7}$ & $2.5\cdot10^{-15}$ & $2.0\cdot10^{- 8}$ \\
\toprule $\mathrm{erf}(z)$ &&&&&& \\ \midrule
time $[$cycles$]$	& $3.2\cdot10^3$ & $5.4\cdot10^1$ & $3.7\cdot10^3$ & $1.4\cdot10^3$ & $8.6\cdot10^2$ & $7.7\cdot10^2$ \\
$\epsilon$		& $1.2\cdot10^{- 9}$ & $8.4\cdot10^{-17}$ & $1.2\cdot10^{- 9}$ & $6.9\cdot10^{- 5}$ & $1.1\cdot10^{-13}$ & $1.4\cdot10^{- 6}$ \\ 
$\epsilon_\mathrm{max}$	& $8.6\cdot10^{- 8}$ & $5.6\cdot10^{-16}$ & $8.6\cdot10^{- 8}$ & $1.3\cdot10^{- 4}$ & $2.3\cdot10^{-12}$ & $6.0\cdot10^{- 6}$ \\
\toprule $\mathrm{erfc}(z)$ &&&&&& \\ \midrule
time $[$cycles$]$	& $3.1\cdot10^3$ & $1.3\cdot10^3$ & $3.5\cdot10^3$ & $1.3\cdot10^3$ & $8.1\cdot10^2$ & $7.5\cdot10^2$ \\
$\epsilon$		& $2.9\cdot10^{-12}$ & $3.3\cdot10^{-16}$ & $2.9\cdot10^{-12}$ & $2.3\cdot10^{- 7}$ & $4.3\cdot10^{-16}$ & $3.7\cdot10^{- 9}$ \\ 
$\epsilon_\mathrm{max}$	& $3.0\cdot10^{-12}$ & $1.2\cdot10^{-15}$ & $3.0\cdot10^{-12}$ & $3.8\cdot10^{- 7}$ & $2.5\cdot10^{-15}$ & $2.0\cdot10^{- 8}$ \\
\bottomrule \end{tabular} \end{center}
\caption{Performance of the various implementations of $w(z)$,
    $\mathrm{erf}(z)$ and $\mathrm{erfc}(z)$ for $(N+1)\cdot1024$ values of $z$ from
    the ``singularity'' region (see text). Time is measured in CPU cycles per
    evaluation (with variations of 5-10\% between different runs of the
    program on the same machine). $\epsilon$ is the average over the
    relative errors of all points tested, $\epsilon_{\mathrm{max}}$ is the
    maximum relative error seen.
\label{tab:perfsing}
}
\end{table}

\subsubsection{Interpretation}
In terms of accuracy, it seems that the {\tt libcerf} implementation and our
code give the best results with relative errors below $10^{-14}$. The
old implementation in {\sc RooFit} and the one in CERNLIB (on which the old
{\sc RooFit} one is based) behave very similarly, and their relative error is
two orders of magnitude larger. The two ``fast'' implementations offer a
relative error of about $10^{-5}$ for the old implementation in {\sc RooFit}
and about $10^{-7}$ for our code.

Concerning the speed of the different algorithms:
The CERNLIB based implementations are slowest. The {\tt libcerf} 
implementation is about a factor 2.4 faster than the CERNLIB
implementation, whereas our full-precision code is about a factor of 3.5
faster.

The fast version of the old {\sc RooFit} code (our implementation) is a factor of 2 (6) 
faster than the original CERNLIB implementation.

% vim: ft=tex:tw=76

\section{Conclusion}
We have presented the calculations needed to obtain analytic expressions
for integrals of the form
\begin{equation}
\frac{1}{\sqrt{2\pi\sigma^2}}
	\int_{t_1}^{t_2}\d{t}\,t^n\,\int_0^{+\infty}\d{t'}\,e^{-(\Gamma -i\Delta m)t'}
	e^{-\frac{(t-t'-\mu)^2}{2\sigma^2}}\,.
\end{equation}
These integrals have an important application in the description of
the time evolution of neutral mesons, which exhibit particle-anitparticle
mixing. There, sine and cosine terms are multiplied by a decaying exponential,
convolved with a Gaussian experimental resolution function, and multiplied
by a polynomial time acceptance function.
Our analytic expressions permit the fast calculation the relevant terms.
We also provide new fast and accurate routines to calculate the Faddeeva function of a
complex argument numerically. This function is needed to evaluate many
of the above integrals. We inlcude the source code in Appendix \ref{sec:code}.

\section*{Acknowledgements}

The authors would like to thank Wouter Hulsbergen and Vladimir Gligorov for
useful discussions and comments on earlier versions of the text.

\appendix
\section{Neutral Meson Mixing in the Presence of Decay Time Resolution, Special Cases\label{sec:specconvolution}}
This section contains the expressions for the special cases mentioned in Section \ref{sec:convolution}.

\subsection{\boldmath Solution for $\min(1/\Gamma,1/\Delta m) \gg \sigma$}
If $\min(1/\Gamma,1/\Delta m) \gg \sigma$, the Gaussian
$G(t-t',\mu,\sigma)$ in Eq.~\ref{eq:mastereq} becomes too narrow to be
observed, and can be replaced by a delta distribution $\delta(t-t'-\mu)$. Thus
Eq.~\ref{eq:mastereq} becomes
\begin{align}
f(t;\Gamma,\Delta m,\sigma,\mu)
&= \int_0^{+\infty}\,\d{t'}\,e^{-(\Gamma -i\Delta m)t'} \delta(t-t'-\mu) \nonumber\\
&= \left\{\begin{array}{l l}
e^{-(\Gamma -i\Delta m)(t-\mu)} & \textrm{for }t\ge\mu \label{eq:mastereqCase2}\\
0 & \textrm{otherwise}
\end{array}\right.\,.
\end{align}
The normalisation integral is
\begin{align}
I_0(t_1,t_2;\Gamma,\Delta m,\sigma,\mu)
&= \int_{t_1}^{t_2} \d{t}\,f(t;\Gamma,\Delta m, \sigma,\mu) \\
&= \int_{t_1}^{t_2} \d{t}\,e^{-(\Gamma -i\Delta m)(t-\mu)} \\
&= \left[
    -\frac{e^{-(\Gamma -i\Delta m)(t-\mu)}}{\Gamma-i\Delta m}
\right]^{\max(t_2,\mu)}_{\max(t_1,\mu)}\,.
\end{align}

\subsection{\boldmath Solution for $\min(\sigma,1/\Delta m) \gg 1/\Gamma$}
In this case, the lifetime is short compared to any other processes, and we
replace $e^{-\Gamma t'}$ by $\delta(t'-1/\Gamma)/\Gamma$ (the delta
distribution is shifted to $1/\Gamma$, the time expectation value of
$e^{-\Gamma t'}$, and scaled to account for the different normalisations). 
Eq.~\ref{eq:mastereq} thus becomes:
\begin{align}
f(t;\Gamma,\Delta m,\sigma,\mu)
&= \frac{1}{\sqrt{2\pi\sigma^2}}\int_0^{+\infty}\d{t'}\,\frac{\delta(t'-1/\Gamma)}{\Gamma} e^{-i\Delta m t'}
	e^{-\frac{(t-t'-\mu)^2}{2\sigma^2}} \nonumber\\
&= \frac{1}{\Gamma}e^{-i\Delta m/\Gamma}\, G(t,\frac{1}{\Gamma}+\mu, \sigma)\,,\label{eq:mastereqCase3}
\end{align}
and the normalisation integral is
\begin{align}
I_0(t_1,t_2;\Gamma,\Delta m,\sigma,\mu)
&= \int_{t_1}^{t_2}\d{t}\, f(t;\Gamma,\Delta m, \sigma,\mu) \\
&= \int_{t_1}^{t_2}\d{t}\, \frac{e^{-i\Delta m/\Gamma}}{\Gamma}\,G(t,\frac{1}{\Gamma}+\mu, \sigma) \\
&= \frac{e^{-i\Delta m/\Gamma}}{2\Gamma} \left[
\mathrm{erf}\left( \frac{t-\frac{1}{\Gamma}-\mu}{\sqrt{2}\sigma} \right)
\right]^{t_2}_{t_1}\,.
\end{align}

\section{Calculating Moments and Including the Effect of an Acceptance Function, Special Cases\label{sec:specaccconvolution}}
This section contains the expressions for the special cases mentioned in Section
\ref{sec:accconvolution}.
\subsection{\boldmath Solution for $\min(1/\Gamma,1/\Delta m) \gg \sigma$}
For $\min(1/\Gamma,1/\Delta m) \gg \sigma$, $f(t;\Gamma,\Delta m,\sigma,\mu)$ simplifies to
\begin{equation}
    f(t;\Gamma,\Delta m,\sigma,\mu) = \left\{\begin{array}{l l}
	    e^{-(\Gamma-i\Delta m)(t-\mu)} & \textrm{for }t\ge\mu \\
	    0 & \textrm{otherwise}
	\end{array}\right.\,.
	\label{eq:mastereqMomentsCase2}
\end{equation}
One is thus interested in the integrals
\begin{equation*}
    I_k(t_1,t_2;\Gamma,\Delta m,\sigma,\mu) = \int_{t_1}^{t_2}\d{t}\,t^k\,e^{-(\Gamma-i\Delta m)(t-\mu)}\,.
\end{equation*}
Abbreviating $u=\Gamma-i\Delta m$, this can be written as
\begin{align*}
    I_k(t_1,t_2;\Gamma,\Delta m,\sigma,\mu) &= e^{u\mu}\,\int_{t_1}^{t_2}\d{t}\,t^k\,e^{-ut} 
    	= e^{u\mu}\,\int_{t_1}^{t_2}\d{t}\,\left.\frac{\d{}^k}{\d{\lambda}^k}\right|_{\lambda=0} e^{(\lambda -u)t} \\
    &= e^{u\mu}\,\left.\frac{\d{}^k}{\d{\lambda}^k}\right|_{\lambda=0}\,\int_{t_1}^{t_2}\d{t}\, e^{(\lambda -u)t} 
    	= e^{u\mu}\,\left.\frac{\d{}^k}{\d{\lambda}^k}\right|_{\lambda=0}\,\frac{[e^{(\lambda -u)t}]_{t_1}^{t_2}}{\lambda-u} \\
    &= e^{u\mu}\,\sum_{j=0}^k{k \choose j}
    	\left(\left.\frac{\d{}^j}{\d{\lambda}^j}\right|_{\lambda=0}\,\frac{1}{\lambda-u}\right)
    	\left.\left(\left.\frac{\d{}^{k-j}}{\d{\lambda}^{k-j}}\right|_{\lambda=0}\,e^{(\lambda-u)t}\right)\right|_{t_1}^{t_2} \\
    &\equiv e^{u\mu}\,\sum_{j=0}^k{k \choose j} G_j(u)\,\left[H_{k-j}(t;u)\right]_{t_1}^{t_2}\,.
\end{align*}
The newly introduced functions $G_n(u)=-n!/u^n$ and $H_n(t;u)=t^n\,e^{-ut}$ are easily computed.
This leaves us with:
\begin{equation*}
    I_k(t_1,t_2;u,\mu) = e^{u\mu}\,\sum_{j=0}^k{k \choose j}
    G_j(u)\,\left[H_{k-j}(t;u)\right]^{\max(\mu,t_2)}_{\max(\mu,t_1)}\,.
\end{equation*}

\subsection{\boldmath Solution for $\min(\sigma,1/\Delta m) \gg 1/\Gamma$}
For $\min(\sigma,1/\Delta m) \gg 1/\Gamma$, Eq.~\ref{eq:mastereq} simplifies to
\begin{equation}
    f(t;\Gamma,\Delta m,\sigma,\mu) =
    \frac{e^{-i\Delta m/\Gamma}}{\Gamma}\cdot \frac{1}{\sqrt{2\pi\sigma^2}} e^{-\frac{(t-1/\Gamma-\mu)^2}{2\sigma^2}}\,,
    \label{eq:mastereqMomentsCase3}
\end{equation}
as shown in Eq.~\ref{eq:mastereqCase3}.
We are again interested in the integrals
\begin{equation}
    I_k(t_1,t_2;\Gamma,\Delta m,\sigma,\mu) =
    \frac{e^{-i\Delta m/\Gamma}}{\Gamma}\cdot
    \int_{\mathrm{max(0,t_1)}}^{\mathrm{max(0,t_2)}}\d{t}\,t^k\,
    \frac{1}{\sqrt{2\pi\sigma^2}} e^{-\frac{(t-1/\Gamma-\mu)^2}{2\sigma^2}}\,.
    \label{eq:integral3}
\end{equation}
Substituting $s=t-\frac{1}{\Gamma}-\mu$ and adjusting the
limits to $s_1=-1/\Gamma-\mu+\max(0,t_1)$ and
$s_2=-1/\Gamma-\mu+\max(0,t_2)$, we obtain
\begin{eqnarray}
    I_k(s_1,s_2;\Gamma,\Delta m,\sigma,\mu) &=&
    \frac{e^{-i\Delta m/\Gamma}}{\Gamma}\cdot
    \int_{s_1}^{s_2}\d{s}\,(s+\frac{1}{\Gamma}+\mu)^k\,
    \frac{1}{\sqrt{2\pi\sigma^2}} e^{-\frac{s^2}{2\sigma^2}} \\
    &=& 
    \frac{e^{-i\Delta m/\Gamma}}{\Gamma}\cdot\frac{1}{\sqrt{2\pi\sigma^2}}\cdot
    \int_{s_1}^{s_2}\d{s}\, \sum_{j=0}^{k}{k \choose j}s^j\left(\frac{1}{\Gamma}+\mu\right)^{k-j}
    e^{-\frac{s^2}{2\sigma^2}} \\
    &=&
    \frac{e^{-i\Delta m/\Gamma}}{\Gamma}\cdot
    \sum_{j=0}^{k}{k \choose j}
    \left(\frac{1}{\Gamma}+\mu\right)^{k-j}
    \frac{1}{\sqrt{2\pi\sigma^2}}\cdot
    \int_{s_1}^{s_2}\d{s}\, s^j e^{-\frac{s^2}{2\sigma^2}} \\
    &\equiv&
    \frac{e^{-i\Delta m/\Gamma}}{\Gamma}\cdot
    \sum_{j=0}^{k}{k \choose j}
    \left(\frac{1}{\Gamma}+\mu\right)^{k-j}
    \cdot I_j(s_1,s_2;\sigma)\,,
\end{eqnarray}
where we have defined
\begin{equation}
I_n(s_1,s_2;\sigma)=\frac{1}{\sqrt{2\pi\sigma^2}}\cdot
\int_{s_1}^{s_2}\d{s}\,s^n e^{-\frac{s^2}{2\sigma^2}}\,.
\end{equation}
The term $s^n$ can be
rewritten in an analogous way as before, completing the square as well
\begin{equation}
    I_n(s_1,s_2;\sigma)=\frac{\sigma^{2n}}{\sqrt{2\pi\sigma^2}}\left.\frac{\d{}^n}{\d{\lambda}^n}\right|_{\lambda=0}
	\int_{s_1}^{s_2}\d{s}\,e^{-\frac{s^2-2s\lambda+\lambda^2-\lambda^2}{2\sigma^2}}\,.
\end{equation}
Substituting $r=\frac{s-\lambda}{\sigma\sqrt{2}}$ yields
\begin{eqnarray}
    I_n(s_1,s_2;\sigma)&=&\frac{\sigma^{2n}}{\sqrt{\pi}}\left.\frac{\d{}^n}{\d{\lambda}^n}\right|_{\lambda=0}
    e^{-\frac{\lambda^2}{2\sigma^2}}\int_{r_1}^{r_2}\d{r}\, e^{-r^2} \\
    &=&\frac{\sigma^{2n}}{2}\left.\frac{\d{}^n}{\d{\lambda}^n}\right|_{\lambda=0}
    e^{-\frac{\lambda^2}{2\sigma^2}}\cdot\mathrm{erf}\left.\left(\frac{s-\lambda}{\sigma\sqrt{2}}\right)\right|_{s_1}^{s_2} \\
    &=& \frac{\sigma^{2n}}{2}\sum_{l=0}^n {n \choose l}
    \left(\left.\frac{\d{}^l}{\d{\lambda}^l}\right|_{\lambda=0}
	e^{-\frac{\lambda^2}{2\sigma^2}} \right)
	\left(\left.\left.\frac{\d{}^{n-l}}{\d{\lambda}^{n-l}}\right|_{\lambda=0}
	\mathrm{erf}\left(\frac{s-\lambda}{\sigma\sqrt{2}}\right)\right|_{s_1}^{s_2}
    \right) \\
    &\equiv& \frac{\sigma^{2n}}{2}\sum_{l=0}^n {n \choose l} P_l(\sigma)\cdot \left. Q_{n-l}(s;\sigma)\right|_{s_1}^{s_2}\,.
\end{eqnarray}
Using the definitions
\begin{equation}
P_n(\sigma)=\left.\frac{\d{}^l}{\d{\lambda}^l}\right|_{\lambda=0}
e^{-\frac{\lambda^2}{2\sigma^2}}\,,\qquad
Q_n(s;\sigma)=\left.\frac{\d{}^{n-l}}{\d{\lambda}^{n-l}}
\right|_{\lambda=0}\mathrm{erf}\left(\frac{s-\lambda}{\sigma\sqrt{2}}\right)\,,
\end{equation}
we tabulate $P_n$ and $Q_n$ for $0\le n\le 3$:
\begin{center}\begin{tabular}{l c c} \toprule
$n$ & $P_n(\sigma)$ & $Q_n(s,\sigma)$ \\ \midrule
    $0$ & $1$ 			& $\mathrm{erf}\left(\frac{s}{\sigma\sqrt{2}}\right)$\\
    $1$ & $0$ 			& $-\frac{\sqrt{2}}{\sqrt{\pi}\sigma}e^{-\frac{s^2}{2\sigma^2}}$ \\
$2$ & $-\frac{1}{\sigma^2}$ 	& $-\frac{\sqrt{2}s}{\sqrt{\pi}\sigma^3}e^{-\frac{s^2}{2\sigma^2}}$ \\ 
    $3$ & $0$ 			& $\frac{\sqrt{2}}{\sqrt{\pi}\sigma^3}e^{-\frac{s^2}{2\sigma^2}}\cdot (1-\frac{s^2}{\sigma^2})$\\
\bottomrule\end{tabular}\end{center}
Eq.~\ref{eq:integral3} can thus be written as:
\begin{eqnarray}
    I_k(t_1,t_2;\Gamma,\Delta m,\sigma,\mu) &=&
    \frac{e^{-i\Delta m/\Gamma}}{2\Gamma}\cdot 
    \sum_{j=0}^{k}{k \choose j} \left(\frac{1}{\Gamma}+\mu\right)^{k-j} \cdot
    \sigma^{2j} \cdot \nonumber\\
    && \qquad\qquad
    \sum_{l=0}^j {j \choose l} P_l(\sigma)\cdot \Big[
    Q_{j-l}(t-\frac{1}{\Gamma}-\mu;\sigma)
    \Big]_{\max(0,t_1)}^{\max(0,t_2)}\,.
\end{eqnarray}

\section{{\tt C++} source code \label{sec:code}}
This section contains the source code of our Faddeeva function implementation.
It has been slightly modified with respect to what is included in ROOT version
5.34/08 to allow standalone builds.

\subsection{File {\tt cerf.h}}
\begin{lstlisting}
#ifndef CERF_H
#define CERF_H

#include <cmath>
#include <complex>

namespace Cerf {
  /** @brief evaluate Faddeeva function for complex argument
   *
   * @author Manuel Schiller <manuel.schiller@nikhef.nl>
   * @date 2013-02-21
   *
   * Calculate the value of the Faddeeva function @f$w(z) = \exp(-z^2)
   * \mathrm{erfc}(-i z)@f$.
   *
   * The method described in
   *
   * S.M. Abrarov, B.M. Quine: "Efficient algorithmic implementation of
   * Voigt/complex error function based on exponential series approximation"
   * published in Applied Mathematics and Computation 218 (2011) 1894-1902
   * doi:10.1016/j.amc.2011.06.072
   *
   * is used. At the heart of the method (equation (14) of the paper) is the
   * following Fourier series based approximation:
   *
   * @f[ w(z) \approx \frac{i}{2\sqrt{\pi}}\left(
   * \sum^N_{n=0} a_n \tau_m\left(
   * \frac{1-e^{i(n\pi+\tau_m z)}}{n\pi + \tau_m z} -
   * \frac{1-e^{i(-n\pi+\tau_m z)}}{n\pi - \tau_m z}
   * \right) - a_0 \frac{1-e^{i \tau_m z}}{z}
   * \right) @f]
   * 
   * The coefficients @f$a_b@f$ are given by:
   *
   * @f[ a_n=\frac{2\sqrt{\pi}}{\tau_m}
   * \exp\left(-\frac{n^2\pi^2}{\tau_m^2}\right) @f]
   *
   * To achieve machine accuracy in double precision floating point arithmetic
   * for most of the upper half of the complex plane, chose @f$t_m=12@f$ and
   * @f$N=23@f$ as is done in the paper.
   *
   * There are two complications: For Im(z) negative, the exponent in the
   * equation above becomes so large that the roundoff in the rest of the
   * calculation is amplified enough that the result cannot be trusted.
   * Therefore, for Im(z) < 0, the symmetry of the erfc function under the
   * transformation z --> -z is used to avoid accuracy issues for Im(z) < 0 by
   * formulating the problem such that the calculation can be done for Im(z) > 0
   * where the accuracy of the method is fine, and some postprocessing then
   * yields the desired final result.
   *
   * Second, the denominators in the equation above become singular at
   * @f$z = n * pi / 12@f$ (for 0 <= n < 24). In a tiny disc around these
   * points, Taylor expansions are used to overcome that difficulty.
   *
   * This routine precomputes everything it can, and tries to write out complex
   * operations to minimise subroutine calls, e.g. for the multiplication of
   * complex numbers.
   *
   * In the square -8 <= Re(z) <= 8, -8 <= Im(z) <= 8, the routine is accurate
   * to better than 4e-13 relative, the average relative error is better than
   * 7e-16. On a modern x86_64 machine, the routine is roughly three times as
   * fast than the old CERNLIB implementation and offers better accuracy.
   */
  std::complex<double> faddeeva(std::complex<double> z);
  /** @brief evaluate Faddeeva function for complex argument (fast version)
   *
   * @author Manuel Schiller <manuel.schiller@nikhef.nl>
   * @date 2013-02-21
   *
   * Calculate the value of the Faddeeva function @f$w(z) = \exp(-z^2)
   * \mathrm{erfc}(-i z)@f$.
   *
   * This is the "fast" version of the faddeeva routine above. Fast means that
   * is takes roughly half the amount of CPU of the slow version of the
   * routine, but is a little less accurate.
   *
   * To be fast, chose @f$t_m=8@f$ and @f$N=11@f$ which should give accuracies
   * around 1e-7.
   *
   * In the square -8 <= Re(z) <= 8, -8 <= Im(z) <= 8, the routine is accurate
   * to better than 4e-7 relative, the average relative error is better than
   * 5e-9. On a modern x86_64 machine, the routine is roughly five times as
   * fast than the old CERNLIB implementation, or about 30% faster than the
   * interpolation/lookup table based fast method used previously in RooFit,
   * and offers better accuracy than the latter (the relative error is roughly
   * a factor 280 smaller than the old interpolation/table lookup routine).
   */
  std::complex<double> faddeeva_fast(std::complex<double> z);

  /** @brief complex erf function
   *
   * @author Manuel Schiller <manuel.schiller@nikhef.nl>
   * @date 2013-02-21
   *
   * Calculate erf(z) for complex z.
   */
  std::complex<double> erf(const std::complex<double> z);

  /** @brief complex erf function (fast version)
   *
   * @author Manuel Schiller <manuel.schiller@nikhef.nl>
   * @date 2013-02-21
   *
   * Calculate erf(z) for complex z. Use the code in faddeeva_fast to save some time.
   */
  std::complex<double> erf_fast(const std::complex<double> z);
  /** @brief complex erfc function
   *
   * @author Manuel Schiller <manuel.schiller@nikhef.nl>
   * @date 2013-02-21
   *
   * Calculate erfc(z) for complex z.
   */
  std::complex<double> erfc(const std::complex<double> z);
  /** @brief complex erfc function (fast version)
   *
   * @author Manuel Schiller <manuel.schiller@nikhef.nl>
   * @date 2013-02-21
   *
   * Calculate erfc(z) for complex z. Use the code in faddeeva_fast to save some time.
   */
  std::complex<double> erfc_fast(const std::complex<double> z);
}

#endif
\end{lstlisting}

\subsection{File {\tt cerf.cc}}
\begin{lstlisting}
#include <complex>
#include <cmath>
#include <algorithm>

#include "cerf.h"

namespace faddeeva_impl {
    static inline void cexp(double& re, double& im)
    {
	// with gcc on unix machines and on x86_64, we can gain by hand-coding
	// exp(z) for the x87 coprocessor; other platforms have the default
	// routines as fallback implementation, and compilers other than gcc on
	// x86_64 generate better code with the default routines; also avoid
	// the inline assembly code when the compiler is not optimising code, or
	// is optimising for code size
	// (we insist on __unix__ here, since the assemblers on other OSs
	// typically do not speak AT&T syntax as gas does...)
#if !defined(__GNUC__) || !defined(__unix__) || !defined(__x86_64__) || \
	!defined(__OPTIMIZE__) || defined(__OPTIMIZE_SIZE__) || \
	defined(__INTEL_COMPILER) || defined(__clang__) || \
	defined(__OPEN64__) || defined(__PATHSCALE__)
	const double e = std::exp(re);
	re = e * std::cos(im);
	im = e * std::sin(im);
#else
	__asm__ (
		"fxam\n\t"		 // examine st(0): NaN? Inf?
		"fstsw %%ax\n\t"
		"movb $0x45,%%dh\n\t"
		"andb %%ah,%%dh\n\t"
		"cmpb $0x05,%%dh\n\t"
		"jz 1f\n\t"		 // have NaN or infinity, handle below
		"fldl2e\n\t"		 // load log2(e)
		"fmulp\n\t"		 // re * log2(e)
		"fld %%st(0)\n\t"	 // duplicate re * log2(e)
		"frndint\n\t"		 // int(re * log2(e))
		"fsubr %%st,%%st(1)\n\t" // st(1) = x = frac(re * log2(e))
		"fxch\n\t"		 // swap st(0), st(1)
		"f2xm1\n\t"		 // 2^x - 1
		"fld1\n\t"		 // st(0) = 1
		"faddp\n\t"		 // st(0) = 2^x
		"fscale\n\t"		 // 2 ^ (int(re * log2(e)) + x)
		"fstp %%st(1)\n\t"	 // pop st(1)
		"jmp 2f\n\t"
		"1:\n\t"		 // handle NaN, Inf...
		"testl $0x200, %%eax\n\t"// -infinity?
		"jz 2f\n\t"
		"fstp %%st\n\t"		 // -Inf, so pop st(0)
		"fldz\n\t"		 // st(0) = 0
		"2:\n\t"		 // here. we have st(0) == exp(re)
		"fxch\n\t"		 // st(0) = im, st(1) = exp(re)
		"fsincos\n\t"		 // st(0) = cos(im), st(1) = sin(im)
		"fnstsw %%ax\n\t"
		"testl $0x400,%%eax\n\t"
		"jz 4f\n\t"		 // |im| too large for fsincos?
		"fldpi\n\t"		 // st(0) = pi
		"fadd %%st(0)\n\t"	 // st(0) *= 2;
		"fxch %%st(1)\n\t"	 // st(0) = im, st(1) = 2 * pi
		"3:\n\t"
		"fprem1\n\t"		 // st(0) = fmod(im, 2 * pi)
		"fnstsw %%ax\n\t"
		"testl $0x400,%%eax\n\t"
		"jnz 3b\n\t"		 // fmod done?
		"fstp %%st(1)\n\t"	 // yes, pop st(1) == 2 * pi
		"fsincos\n\t"		 // st(0) = cos(im), st(1) = sin(im)
		"4:\n\t"		 // all fine, fsincos succeeded
		"fmul %%st(2)\n\t"	 // st(0) *= st(2)
		"fxch %%st(2)\n\t"	 // st(2)=exp(re)*cos(im),st(0)=exp(im)
		"fmulp %%st(1)\n\t"	 // st(1)=exp(re)*sin(im), pop st(0)
		: "=t" (im), "=u" (re): "0" (re), "1" (im) :
		    "eax", "dh", "cc", "st(5)", "st(6)", "st(7)");
#endif
    }

    template <class T, unsigned N, unsigned NTAYLOR>
    static inline std::complex<T> faddeeva_smabmq_impl(
	    T zre, T zim, const T tm,
	    const T (&a)[N], const T (&npi)[N],
	    const T (&taylorarr)[N * NTAYLOR * 2])
    {
	// catch singularities in the Fourier representation At
	// z = n pi / tm, and provide a Taylor series expansion in those
	// points, and only use it when we're close enough to the real axis
	// that there is a chance we need it
	const T zim2 = zim * zim;
	const T maxnorm = T(9) / T(1000000);
	if (zim2 < maxnorm) {
	    // we're close enough to the real axis that we need to worry about
	    // singularities
	    const T dnsing = tm * zre / npi[1];
	    const T dnsingmax2 = (T(N) - T(1) / T(2)) * (T(N) - T(1) / T(2));
	    if (dnsing * dnsing < dnsingmax2) {
		// we're in the interesting range of the real axis as well...
		// deal with Re(z) < 0 so we only need N different Taylor
		// expansions; use w(-x+iy) = conj(w(x+iy))
		const bool negrez = zre < 0.;
		// figure out closest singularity
		const int nsing = int(std::abs(dnsing) + T(1) / T(2));
		// and calculate just how far we are from it
		const T zmnpire = std::abs(zre) - npi[nsing];
		const T zmnpinorm = zmnpire * zmnpire + zim2;
		// close enough to one of the singularities?
		if (zmnpinorm < maxnorm) {
		    const T* coeffs = &taylorarr[nsing * NTAYLOR * 2];
		    // calculate value of taylor expansion...
		    // (note: there's no chance to vectorize this one, since
		    // the value of the next iteration depend on the ones from
		    // the previous iteration)
		    T sumre = coeffs[0], sumim = coeffs[1];
		    for (unsigned i = 1; i < NTAYLOR; ++i) {
			const T re = sumre * zmnpire - sumim * zim;
			const T im = sumim * zmnpire + sumre * zim;
			sumre = re + coeffs[2 * i + 0];
			sumim = im + coeffs[2 * i + 1];
		    }
		    // undo the flip in real part of z if needed
		    if (negrez) return std::complex<T>(sumre, -sumim);
		    else return std::complex<T>(sumre, sumim);
		}
	    }
	}
	// negative Im(z) is treated by calculating for -z, and using the
	// symmetry properties of erfc(z)
	const bool negimz = zim < 0.;
	if (negimz) {
	    zre = -zre;
	    zim = -zim;
	}
	const T twosqrtpi = 3.54490770181103205e+00;
	const T tmzre = tm * zre, tmzim = tm * zim;
	// calculate exp(i tm z)
	T eitmzre = -tmzim, eitmzim = tmzre;
	faddeeva_impl::cexp(eitmzre, eitmzim);
	// form 1 +/- exp (i tm z)
	const T numerarr[4] = {
	    T(1) - eitmzre, -eitmzim, T(1) + eitmzre, +eitmzim
	};
	// form tm z * (1 +/- exp(i tm z))
	const T numertmz[4] = {
	    tmzre * numerarr[0] - tmzim * numerarr[1],
	    tmzre * numerarr[1] + tmzim * numerarr[0],
	    tmzre * numerarr[2] - tmzim * numerarr[3],
	    tmzre * numerarr[3] + tmzim * numerarr[2]
	};
	// common subexpressions for use inside the loop
	const T reimtmzm2 = T(-2) * tmzre * tmzim;
	const T imtmz2 = tmzim * tmzim;
	const T reimtmzm22 = reimtmzm2 * reimtmzm2;
	// on non-x86_64 architectures, when the compiler is producing
	// unoptimised code and when optimising for code size, we use the
	// straightforward implementation, but for x86_64, we use the
	// brainf*cked code below that the gcc vectorizer likes to gain a few
	// clock cycles; non-gcc compilers also get the normal code, since they
	// usually do a better job with the default code (and yes, it's a pain
	// that they're all pretending to be gcc)
#if (!defined(__x86_64__)) || !defined(__OPTIMIZE__) || \
	defined(__OPTIMIZE_SIZE__) || defined(__INTEL_COMPILER) || \
	defined(__clang__) || defined(__OPEN64__) || \
	defined(__PATHSCALE__) || !defined(__GNUC__)
        const T znorm = zre * zre + zim2;
        T sumre = (-a[0] / znorm) * (numerarr[0] * zre + numerarr[1] * zim);
        T sumim = (-a[0] / znorm) * (numerarr[1] * zre - numerarr[0] * zim);
        for (unsigned i = 0; i < N; ++i) {
            const unsigned j = (i << 1) & 2;
            // denominator
            const T wk = imtmz2 + (npi[i] + tmzre) * (npi[i] - tmzre);
            // norm of denominator
            const T norm = wk * wk + reimtmzm22;
            const T f = T(2) * tm * a[i] / norm;
            // sum += a[i] * numer / wk
            sumre -= f * (numertmz[j] * wk + numertmz[j + 1] * reimtmzm2);
            sumim -= f * (numertmz[j + 1] * wk - numertmz[j] * reimtmzm2);
        }
#else
	// BEGIN fully vectorisable code - enjoy reading... ;)
	T tmp[2 * N];
	for (unsigned i = 0; i < N; ++i) {
	    const T wk = imtmz2 + (npi[i] + tmzre) * (npi[i] - tmzre);
	    tmp[2 * i + 0] = wk;
	    tmp[2 * i + 1] = T(2) * tm * a[i] / (wk * wk + reimtmzm22);
	}
	for (unsigned i = 0; i < N / 2; ++i) {
	    T wk = tmp[4 * i + 0], f = tmp[4 * i + 1];
	    tmp[4 * i + 0] = -f * (numertmz[0] * wk + numertmz[1] * reimtmzm2);
	    tmp[4 * i + 1] = -f * (numertmz[1] * wk - numertmz[0] * reimtmzm2);
	    wk = tmp[4 * i + 2], f = tmp[4 * i + 3];
	    tmp[4 * i + 2] = -f * (numertmz[2] * wk + numertmz[3] * reimtmzm2);
	    tmp[4 * i + 3] = -f * (numertmz[3] * wk - numertmz[2] * reimtmzm2);
	}
	if (N & 1) {
	    // we may have missed one element in the last loop; if so, process
	    // it now...
	    const T wk = tmp[2 * N - 2], f = tmp[2 * N - 1];
	    tmp[2 * (N - 1) + 0] = -f * (numertmz[0] * wk + numertmz[1] * reimtmzm2);
	    tmp[2 * (N - 1) + 1] = -f * (numertmz[1] * wk - numertmz[0] * reimtmzm2);
	}
	const T znorm = zre * zre + zim2;
	T sumre = (-a[0] / znorm) * (numerarr[0] * zre + numerarr[1] * zim);
	T sumim = (-a[0] / znorm) * (numerarr[1] * zre - numerarr[0] * zim);
	for (unsigned i = 0; i < N; ++i) {
	    sumre += tmp[2 * i + 0];
	    sumim += tmp[2 * i + 1];
	}
	// END fully vectorisable code
#endif
	// prepare the result
	if (negimz) {
	    // use erfc(-z) = 2 - erfc(z) to get good accuracy for
	    // Im(z) < 0: 2 / exp(z^2) - w(z)
	    const T z2im = T(2) * zre * zim;
	    const T z2re = (zre + zim) * (zre - zim);
	    T ez2re = z2re, ez2im = z2im;
	    faddeeva_impl::cexp(ez2re, ez2im);
	    const T twoez2norm = T(2) / (ez2re * ez2re + ez2im * ez2im);
	    return std::complex<T>(twoez2norm * ez2re + sumim / twosqrtpi,
		    -twoez2norm * ez2im - sumre / twosqrtpi);
	} else {
	    return std::complex<T>(-sumim / twosqrtpi, sumre / twosqrtpi);
	}
    }

    static const double npi24[24] = { // precomputed values n * pi
	0.00000000000000000e+00, 3.14159265358979324e+00, 6.28318530717958648e+00,
	9.42477796076937972e+00, 1.25663706143591730e+01, 1.57079632679489662e+01,
	1.88495559215387594e+01, 2.19911485751285527e+01, 2.51327412287183459e+01,
	2.82743338823081391e+01, 3.14159265358979324e+01, 3.45575191894877256e+01,
	3.76991118430775189e+01, 4.08407044966673121e+01, 4.39822971502571053e+01,
	4.71238898038468986e+01, 5.02654824574366918e+01, 5.34070751110264851e+01,
	5.65486677646162783e+01, 5.96902604182060715e+01, 6.28318530717958648e+01,
	6.59734457253856580e+01, 6.91150383789754512e+01, 7.22566310325652445e+01,
    };
    static const double a24[24] = { // precomputed Fourier coefficient prefactors
	2.95408975150919338e-01, 2.75840233292177084e-01, 2.24573955224615866e-01,
	1.59414938273911723e-01, 9.86657664154541891e-02, 5.32441407876394120e-02,
	2.50521500053936484e-02, 1.02774656705395362e-02, 3.67616433284484706e-03,
	1.14649364124223317e-03, 3.11757015046197600e-04, 7.39143342960301488e-05,
	1.52794934280083635e-05, 2.75395660822107093e-06, 4.32785878190124505e-07,
	5.93003040874588103e-08, 7.08449030774820423e-09, 7.37952063581678038e-10,
	6.70217160600200763e-11, 5.30726516347079017e-12, 3.66432411346763916e-13,
	2.20589494494103134e-14, 1.15782686262855879e-15, 5.29871142946730482e-17,
    };
    static const double taylorarr24[24 * 12] = {
	// real part imaginary part, low order coefficients last
	// nsing = 0
	 0.00000000000000000e-00,  3.00901111225470020e-01,
	 5.00000000000000000e-01,  0.00000000000000000e-00,
	 0.00000000000000000e-00, -7.52252778063675049e-01,
	-1.00000000000000000e-00,  0.00000000000000000e-00,
	 0.00000000000000000e-00,  1.12837916709551257e+00,
	 1.00000000000000000e-00,  0.00000000000000000e-00,
	// nsing = 1
	-2.22423508493755319e-01,  1.87966717746229718e-01,
	 3.41805419240637628e-01,  3.42752593807919263e-01,
	 4.66574321730757753e-01, -5.59649213591058097e-01,
	-8.05759710273191021e-01, -5.38989366115424093e-01,
	-4.88914083733395200e-01,  9.80580906465856792e-01,
	 9.33757118080975970e-01,  2.82273885115127769e-01,
	// nsing = 2
	-2.60522586513312894e-01, -4.26259455096092786e-02,
	 1.36549702008863349e-03,  4.39243227763478846e-01,
	 6.50591493715480700e-01, -1.23422352472779046e-01,
	-3.43379903564271318e-01, -8.13862662890748911e-01,
	-7.96093943501906645e-01,  6.11271022503935772e-01,
	 7.60213717643090957e-01,  4.93801903948967945e-01,
	// nsing = 3
	-1.18249853727020186e-01, -1.90471659765411376e-01,
	-2.59044664869706839e-01,  2.69333898502392004e-01,
	 4.99077838344125714e-01,  2.64644800189075006e-01,
	 1.26114512111568737e-01, -7.46519337025968199e-01,
	-8.47666863706379907e-01,  1.89347715957263646e-01,
	 5.39641485816297176e-01,  5.97805988669631615e-01,
	// nsing = 4
	 4.94825297066481491e-02, -1.71428212158876197e-01,
	-2.97766677111471585e-01,  1.60773286596649656e-02,
	 1.88114210832460682e-01,  4.11734391195006462e-01,
	 3.98540613293909842e-01, -4.63321903522162715e-01,
	-6.99522070542463639e-01, -1.32412024008354582e-01,
	 3.33997185986131785e-01,  6.01983450812696742e-01,
	// nsing = 5
	 1.18367078448232332e-01, -6.09533063579086850e-02,
	-1.74762998833038991e-01, -1.39098099222000187e-01,
	-6.71534655984154549e-02,  3.34462251996496680e-01,
	 4.37429678577360024e-01, -1.59613865629038012e-01,
	-4.71863911886034656e-01, -2.92759316465055762e-01,
	 1.80238737704018306e-01,  5.42834914744283253e-01,
	// nsing = 6
	 8.87698096005701290e-02,  2.84339354980994902e-02,
	-3.18943083830766399e-02, -1.53946887977045862e-01,
	-1.71825061547624858e-01,  1.70734367410600348e-01,
	 3.33690792296469441e-01,  3.97048587678703930e-02,
	-2.66422678503135697e-01, -3.18469797424381480e-01,
	 8.48049724711137773e-02,  4.60546329221462864e-01,
	// nsing = 7
	 2.99767046276705077e-02,  5.34659695701718247e-02,
	 4.53131030251822568e-02, -9.37915401977138648e-02,
	-1.57982359988083777e-01,  3.82170507060760740e-02,
	 1.98891589845251706e-01,  1.17546677047049354e-01,
	-1.27514335237079297e-01, -2.72741112680307074e-01,
	 3.47906344595283767e-02,  3.82277517244493224e-01,
	// nsing = 8
	-7.35922494437203395e-03,  3.72011290318534610e-02,
	 5.66783220847204687e-02, -3.21015398169199501e-02,
	-1.00308737825172555e-01, -2.57695148077963515e-02,
	 9.67294850588435368e-02,  1.18174625238337507e-01,
	-5.21266530264988508e-02, -2.08850084114630861e-01,
	 1.24443217440050976e-02,  3.19239968065752286e-01,
	// nsing = 9
	-1.66126772808035320e-02,  1.46180329587665321e-02,
	 3.85927576915247303e-02,  1.18910471133003227e-03,
	-4.94003498320899806e-02, -3.93468443660139110e-02,
	 3.92113167048952835e-02,  9.03306084789976219e-02,
	-1.82889636251263500e-02, -1.53816215444915245e-01,
	 3.88103861995563741e-03,  2.72090310854550347e-01,
	// nsing = 10
	-1.21245068916826880e-02,  1.59080224420074489e-03,
	 1.91116222508366035e-02,  1.05879549199053302e-02,
	-1.97228428219695318e-02, -3.16962067712639397e-02,
	 1.34110372628315158e-02,  6.18045654429108837e-02,
	-5.52574921865441838e-03, -1.14259663804569455e-01,
	 1.05534036292203489e-03,  2.37326534898818288e-01,
	// nsing = 11
	-5.96835002183177493e-03, -2.42594931567031205e-03,
	 7.44753817476594184e-03,  9.33450807578394386e-03,
	-6.52649522783026481e-03, -2.08165802069352019e-02,
	 3.89988065678848650e-03,  4.12784313451549132e-02,
	-1.44110721106127920e-03, -8.76484782997757425e-02,
	 2.50210184908121337e-04,  2.11131066219336647e-01,
	// nsing = 12
	-2.24505212235034193e-03, -2.38114524227619446e-03,
	 2.36375918970809340e-03,  5.97324040603806266e-03,
	-1.81333819936645381e-03, -1.28126250720444051e-02,
	 9.69251586187208358e-04,  2.83055679874589732e-02,
	-3.24986363596307374e-04, -6.97056268370209313e-02,
	 5.17231862038123061e-05,  1.90681117197597520e-01,
	// nsing = 13
	-6.76887607549779069e-04, -1.48589685249767064e-03,
	 6.22548369472046953e-04,  3.43871156746448680e-03,
	-4.26557147166379929e-04, -7.98854145009655400e-03,
	 2.06644460919535524e-04,  2.03107152586353217e-02,
	-6.34563929410856987e-05, -5.71425144910115832e-02,
	 9.32252179140502456e-06,  1.74167663785025829e-01,
	// nsing = 14
	-1.67596437777156162e-04, -8.05384193869903178e-04,
	 1.37627277777023791e-04,  1.97652692602724093e-03,
	-8.54392244879459717e-05, -5.23088906415977167e-03,
	 3.78965577556493513e-05,  1.52191559129376333e-02,
	-1.07393019498185646e-05, -4.79347862153366295e-02,
	 1.46503970628861795e-06,  1.60471011683477685e-01,
	// nsing = 15
	-3.45715760630978778e-05, -4.31089554210205493e-04,
	 2.57350138106549737e-05,  1.19449262097417514e-03,
	-1.46322227517372253e-05, -3.61303766799909378e-03,
	 5.99057675687392260e-06,  1.17993805017130890e-02,
	-1.57660578509526722e-06, -4.09165023743669707e-02,
	 2.00739683204152177e-07,  1.48879348585662670e-01,
	// nsing = 16
	-5.99735188857573424e-06, -2.42949218855805052e-04,
	 4.09249090936269722e-06,  7.67400152727128171e-04,
	-2.14920611287648034e-06, -2.60710519575546230e-03,
	 8.17591694958640978e-07,  9.38581640137393053e-03,
	-2.00910914042737743e-07, -3.54045580123653803e-02,
	 2.39819738182594508e-08,  1.38916449405613711e-01,
	// nsing = 17
	-8.80708505155966658e-07, -1.46479474515521504e-04,
	 5.55693207391871904e-07,  5.19165587844615415e-04,
	-2.71391142598826750e-07, -1.94439427580099576e-03,
	 9.64641799864928425e-08,  7.61536975207357980e-03,
	-2.22357616069432967e-08, -3.09762939485679078e-02,
	 2.49806920458212581e-09,  1.30247401712293206e-01,
	// nsing = 18
	-1.10007111030476390e-07, -9.35886150886691786e-05,
	 6.46244096997824390e-08,  3.65267193418479043e-04,
	-2.95175785569292542e-08, -1.48730955943961081e-03,
	 9.84949251974795537e-09,  6.27824679148707177e-03,
	-2.13827217704781576e-09, -2.73545766571797965e-02,
	 2.26877724435352177e-10,  1.22627158810895267e-01,
	// nsing = 19
	-1.17302439957657553e-08, -6.24890956722053332e-05,
	 6.45231881609786173e-09,  2.64799907072561543e-04,
	-2.76943921343331654e-09, -1.16094187847598385e-03,
	 8.71074689656480749e-10,  5.24514377390761210e-03,
	-1.78730768958639407e-10, -2.43489203319091538e-02,
	 1.79658223341365988e-11,  1.15870972518909888e-01,
	// nsing = 20
	-1.07084502471985403e-09, -4.31515421260633319e-05,
	 5.54152563270547927e-10,  1.96606443937168357e-04,
	-2.24423474431542338e-10, -9.21550077887211094e-04,
	 6.67734377376211580e-11,  4.43201203646827019e-03,
	-1.29896907717633162e-11, -2.18236356404862774e-02,
	 1.24042409733678516e-12,  1.09836276968151848e-01,
	// nsing = 21
	-8.38816525569060600e-11, -3.06091807093959821e-05,
	 4.10033961556230842e-11,  1.48895624771753491e-04,
	-1.57238128435253905e-11, -7.42073499862065649e-04,
	 4.43938379112418832e-12,  3.78197089773957382e-03,
	-8.21067867869285873e-13, -1.96793607299577220e-02,
	 7.46725770201828754e-14,  1.04410965521273064e-01,
	// nsing = 22
	-5.64848922712870507e-12, -2.22021942382507691e-05,
	 2.61729537775838587e-12,  1.14683068921649992e-04,
	-9.53316139085394895e-13, -6.05021573565916914e-04,
	 2.56116039498542220e-13,  3.25530796858307225e-03,
	-4.51482829896525004e-14, -1.78416955716514289e-02,
	 3.91940313268087086e-15,  9.95054815464739996e-02,
	// nsing = 23
	-3.27482357793897640e-13, -1.64138890390689871e-05,
	 1.44278798346454523e-13,  8.96362542918265398e-05,
	-5.00524303437266481e-14, -4.98699756861136127e-04,
	 1.28274026095767213e-14,  2.82359118537843949e-03,
	-2.16009593993917109e-15, -1.62538825704327487e-02,
	 1.79368667683853708e-16,  9.50473084594884184e-02
    };

    const double npi11[11] = { // precomputed values n * pi
	0.00000000000000000e+00, 3.14159265358979324e+00, 6.28318530717958648e+00,
	9.42477796076937972e+00, 1.25663706143591730e+01, 1.57079632679489662e+01,
	1.88495559215387594e+01, 2.19911485751285527e+01, 2.51327412287183459e+01,
	2.82743338823081391e+01, 3.14159265358979324e+01
    };
    const double a11[11] = { // precomputed Fourier coefficient prefactors
	4.43113462726379007e-01, 3.79788034073635143e-01, 2.39122407410867584e-01,
	1.10599187402169792e-01, 3.75782250080904725e-02, 9.37936104296856288e-03,
	1.71974046186334976e-03, 2.31635559000523461e-04, 2.29192401420125452e-05,
	1.66589592139340077e-06, 8.89504561311882155e-08
    };
    const double taylorarr11[11 * 6] = {
	// real part imaginary part, low order coefficients last
	// nsing = 0
	-1.00000000000000000e+00,  0.00000000000000000e+00,
	 0.00000000000000000e-01,  1.12837916709551257e+00,
	 1.00000000000000000e+00,  0.00000000000000000e+00,
	// nsing = 1
	-5.92741768247463996e-01, -7.19914991991294310e-01,
	-6.73156763521649944e-01,  8.14025039279059577e-01,
	 8.57089811121701143e-01,  4.00248106586639754e-01,
	// nsing = 2
	 1.26114512111568737e-01, -7.46519337025968199e-01,
	-8.47666863706379907e-01,  1.89347715957263646e-01,
	 5.39641485816297176e-01,  5.97805988669631615e-01,
	// nsing = 3
	 4.43238482668529408e-01, -3.03563167310638372e-01,
	-5.88095866853990048e-01, -2.32638360700858412e-01,
	 2.49595637924601714e-01,  5.77633779156009340e-01,
	// nsing = 4
	 3.33690792296469441e-01,  3.97048587678703930e-02,
	-2.66422678503135697e-01, -3.18469797424381480e-01,
	 8.48049724711137773e-02,  4.60546329221462864e-01,
	// nsing = 5
	 1.42043544696751869e-01,  1.24094227867032671e-01,
	-8.31224229982140323e-02, -2.40766729258442100e-01,
	 2.11669512031059302e-02,  3.48650139549945097e-01,
	// nsing = 6
	 3.92113167048952835e-02,  9.03306084789976219e-02,
	-1.82889636251263500e-02, -1.53816215444915245e-01,
	 3.88103861995563741e-03,  2.72090310854550347e-01,
	// nsing = 7
	 7.37741897722738503e-03,  5.04625223970221539e-02,
	-2.87394336989990770e-03, -9.96122819257496929e-02,
	 5.22745478269428248e-04,  2.23361039070072101e-01,
	// nsing = 8
	 9.69251586187208358e-04,  2.83055679874589732e-02,
	-3.24986363596307374e-04, -6.97056268370209313e-02,
	 5.17231862038123061e-05,  1.90681117197597520e-01,
	// nsing = 9
	 9.01625563468897100e-05,  1.74961124275657019e-02,
	-2.65745127697337342e-05, -5.22070356354932341e-02,
	 3.75952450449939411e-06,  1.67018782142871146e-01,
	// nsing = 10
	 5.99057675687392260e-06,  1.17993805017130890e-02,
	-1.57660578509526722e-06, -4.09165023743669707e-02,
	 2.00739683204152177e-07,  1.48879348585662670e-01
    };
}

std::complex<double> Cerf::faddeeva(std::complex<double> z)
{
    return faddeeva_impl::faddeeva_smabmq_impl<double, 24, 6>(
	    z.real(), z.imag(), 12., faddeeva_impl::a24,
	    faddeeva_impl::npi24, faddeeva_impl::taylorarr24);
}

std::complex<double> Cerf::faddeeva_fast(std::complex<double> z)
{
    return faddeeva_impl::faddeeva_smabmq_impl<double, 11, 3>(
	    z.real(), z.imag(), 8., faddeeva_impl::a11,
	    faddeeva_impl::npi11, faddeeva_impl::taylorarr11);
}

std::complex<double> Cerf::erfc(const std::complex<double> z)
{
    double re = -z.real() * z.real() + z.imag() * z.imag();
    double im = -2. * z.real() * z.imag();
    faddeeva_impl::cexp(re, im);
    return (z.real() >= 0.) ?
	(std::complex<double>(re, im) *
	 faddeeva(std::complex<double>(-z.imag(), z.real()))) :
	(2. - std::complex<double>(re, im) *
	 faddeeva(std::complex<double>(z.imag(), -z.real())));
}

std::complex<double> Cerf::erfc_fast(const std::complex<double> z)
{
    double re = -z.real() * z.real() + z.imag() * z.imag();
    double im = -2. * z.real() * z.imag();
    faddeeva_impl::cexp(re, im);
    return (z.real() >= 0.) ?
	(std::complex<double>(re, im) *
	 faddeeva_fast(std::complex<double>(-z.imag(), z.real()))) :
	(2. - std::complex<double>(re, im) *
	 faddeeva_fast(std::complex<double>(z.imag(), -z.real())));
}

std::complex<double> Cerf::erf(const std::complex<double> z)
{
    double re = -z.real() * z.real() + z.imag() * z.imag();
    double im = -2. * z.real() * z.imag();
    faddeeva_impl::cexp(re, im);
    return (z.real() >= 0.) ?
	(1. - std::complex<double>(re, im) *
	 faddeeva(std::complex<double>(-z.imag(), z.real()))) :
	(std::complex<double>(re, im) *
	 faddeeva(std::complex<double>(z.imag(), -z.real())) - 1.);
}

std::complex<double> Cerf::erf_fast(const std::complex<double> z)
{
    double re = -z.real() * z.real() + z.imag() * z.imag();
    double im = -2. * z.real() * z.imag();
    faddeeva_impl::cexp(re, im);
    return (z.real() >= 0.) ?
	(1. - std::complex<double>(re, im) *
	 faddeeva_fast(std::complex<double>(-z.imag(), z.real()))) :
	(std::complex<double>(re, im) *
	 faddeeva_fast(std::complex<double>(z.imag(), -z.real())) - 1.);
}

\end{lstlisting}

\printbibliography
\addcontentsline{toc}{section}{References}

\end{document}